\documentclass[twocolumn,eqsecnum,showkeys,showpacs,aps,prb]{revtex4}

\usepackage{graphicx}

\begin{document}

\title{Spin Splitting in Quantum Wires with Perturbed Axial Symmetry}%

\author{A.D.~Bochevarov${}^\dagger$ and K.~Ilyenko}
% \thanks{Also at the Faculty of Physics, Kharkiv National University.}%
%\email{arteum@ire.kharkov.ua}%
\affiliation{Institute for Radiophysics and Electronics of NASU,
             12 Ak.~Proskura Street, Kharkiv 61085, Ukraine}%

\date{26 November 2001}%

\begin{abstract}
We investigate theoretically perturbations to the confining
potential capable of lifting spin degeneracy in axially symmetric
quasi-one-dimensional electron gases with the spin-orbit
interaction. The  role  of  two  different types of perturbations
breaking axial symmetry of the system: of non-electromagnetic and
electromagnetic origin is analyzed. We found that these two types
of perturbations have a fundamental distinction in their effects
on energy spectra. The influence of the scale of perturbation upon
the value of spin splitting is investigated and it is shown that
under certain conditions such splittings can be observed in the
experiment.
\end{abstract}

%Use showkeys class option if keyword display desired
\maketitle
\section{Introduction}
Spin splitting in energy bands of quasi-one-dimensional
and\hspace{1ex} two-dimensional\hspace{1ex} electron\hspace{2ex}
gases\hspace{1ex} (Q1DEG's\hspace{1ex} and \\
2DEG's, respectively) in zero magnetic field has been studied
extensively both theoretically
\cite{Rashba2,Cardona,Lommer,Das,Moroz} and
ex\-pe\-ri\-men\-tal\-ly \cite{Thornton,Chen,Pikus,Nitta,Bennett}
over the two past decades. Such an interest was partially produced
by the prospect of creating new electronic devices which exploit
spin-polarized transport \cite{Datta,Prinz}. The appearance of
spin splitting in energy spectra of realistic heterostructures,
which are usually described by Q1DEG's and 2DEG's models, was
ascribed mainly to Rashba \cite{Rashba1} and Dresselhaus
\cite{Dresselhaus} terms in the corresponding Hamiltonians. Such
Hamiltonians usually possess two distinctive features: (i) as
parameters varied in theoretical constructs they contain merely
coefficients placed before splitting-inducing terms in order to
achieve an agreement with the experiment \cite{Das} and (ii) they
include confining potentials for effective reduction of the
dimensions of the system. While a thorough analysis of the effect
of different splitting-inducing terms upon energy spectra has been
carried out and their role understood, the choice of confining
potentials is still limited mainly to simple (and model) hard-wall
\cite{MacKinnon} and parabolic \cite{Luax,Drexler,Brataas,Moroz}
ones. Such a choice is conditioned by the fact that the
corresponding non-perturbed problems are often explicitly
solvable. At the same time this may give rather simplified picture
of real energy band structure of the model systems. The common
step to avoid this shortcoming is to change the form of a
potential slightly, for example, by accounting for
non-parabolicity terms. Here, however, we do not touch upon this
and related issues which are unlikely to seriously influence the
general spectral picture, but introduce confining potentials which
are themselves capable of giving rise to spin splitting beyond
Rashba and Dresselhaus mechanisms.

In this paper we investigate an infinite Q1DEG produced out of
3-dimensional system by means of a 2-dimensional confining
potential. The initial confining potential is assumed to be
axially symmetrical and have the electromagnetic nature. Axially
symmetrical quantum wires are rather easy to investigate
theoretically and, what is more important, they are not just a
theoretical model, but may be prepared practically (see, for
instance~\cite{Tonucci}, and references therein). We also neglect
Coulomb interaction between electrons, which is quite realistic
given that it can be accounted for by the renormalization of the
SO coupling constant \cite{Chen2}. Spin-orbit (SO) interaction
emanates from the Pauli equation and is of the form:
\begin{equation}
\frac{e \hbar}{4
m_{e}^{2}c^{2}}\vec{\nabla}\varphi\cdot(\vec{\sigma}\times
\hat{\vec{p}}),
\end{equation}
where $m_{e}$ is the effective electron mass, $\varphi$ is the
scalar electromagnetic potential, $\vec{\sigma}$ is the Pauli
matrices vector, and $\hat{\vec{p}}$ is the momentum operator. The
vector electromagnetic potential $\vec{A}$ is put equal to zero in
all the systems this paper deals with.

The described system possesses a spectrum in which every energy
level is twice degenerate. We introduce a perturbation (not
necessarily small) to the confining potential that breaks axial
symmetry of the system. This leads to lifting spin degeneracy, the
mechanism of which being different depending on whether the
perturbation has electromagnetic or non-electromagnetic origin.
The investigation of effects, which such perturbations have on
spectra, is yet more important if one takes into account that
perturbations conditioning deviations from axial symmetry are
always present in the experiment. We show that under certain
conditions the spin splitting, originated by potentials in
question, may be large enough to be detected.
%%%%%%%%%%%%%%%%%%%%%%%%%%%%%%%%%%%%%%%%%%%%%%%%%%%%%%%%%%%%%%%%%%%%%%%%%%
\section{Axially symmetric confining potential}
Let us first consider the case of axially symmetric confining
potential created by purely electrostatic field. The inclusion of
SO interaction into the Hamiltonian significantly hinders seeking
the solution of the corresponding eigenvalue problem, so that one
may expect to find solutions only by means of perturbation theory
or numerical calculations. That is why we chose a parabolic
confining potential
\begin{equation}
\hat{V}_{conf}=\frac{m_{e}\omega^{2}}{2}\rho^{2},
\end{equation}%
where $\omega$ has the dimension of frequency and measures the
strength of the potential and $\rho$ is the lateral cylindrical
coordinate. This potential will provide us with a convenient zero
approximation to the solution. The Hamiltonian of our problem has
the following form:
\begin{equation}
\hat{H}=\frac{\hat{\vec{p}}^{\;2}}{2m_{e}}+\hat{V}_{conf}+\mu
\hat{V}_{SO} + \frac{\hbar^{2}\omega^{2}}{4m_{e}c^{2}}.
\end{equation}
Here the SO interaction is
\begin{equation}
\hat{V}_{SO}=-\frac{i\hbar^{2}\omega^{2}}{4m_{e}c^{2}}\left[-i\sigma_{z}\frac{\partial}
{\partial\theta}+\rho\left(
\begin{array}{cc}
0 & -i e^{-i\theta} \\ i e^{i\theta} & 0
\end{array}\right)\frac{\partial}{\partial z} \right].
\end{equation}
The positive constant $\mu$
%, analogous to the constant $\beta$
%introduced in \cite{Moroz and Ko}
defines the strength of the SO interaction. At $\mu=1$ the
Hamiltonian (3) naturally stems from the usual Pauli Hamiltonian
$$\hat{H}_{P}=\frac{\hat{\vec{p}}^{\;2}}{2m_{e}}+e\varphi-\frac{\hat{\vec{p}}^{\;4}}{8m_{e}^{3}c^{2}}$$
\begin{equation}
-\frac{e\hbar}{4m_{e}^{2}c^{2}}\vec{\sigma}\cdot(\vec{E}\times\hat{\vec{p}})-\frac{e
\hbar^{2}}{8m_{e}^{2}c^{2}}\mathrm{div} \vec{E}
\end{equation}
if one puts the scalar potential $\varphi$ equal to
$m_{e}\omega^{2}\rho^{2}/2e$ and neglects the term
$\hat{\vec{p}}^{\;4}/8m_{e}^{3}c^{2}$, which does not influence
the spin-splitting.

The solution to the zero approximation eigenvalue problem
\begin{equation}
(\frac{\hat{\vec{p}}^{\;2}}{2m_{e}}+\hat{V}_{conf})\chi\equiv
\hat{H}_{0}\chi=E\chi
\end{equation}
was found in 1928 by Fock \cite{Fock}. It reads
\begin{equation}
\chi_{nlk_{z}}(\rho,z,\theta)=\Psi_{nl}(\rho)Y_{l}(\theta)R_{k_{z}}(z),
\end{equation}
$$\Psi_{nl}(\rho)=\sqrt{\frac{2n!}{(n+|l|)!}}\frac{1}{\rho_{\omega}}\exp
[-\frac{1}{2}\left(\frac{\rho}
{\rho_{\omega}}\right)^{2}]\left(\frac{\rho}{\rho_{\omega}}\right)^{|l|}$$
\begin{equation}
\times L_{n}^{|l|}
\left(\frac{\rho^{2}}{\rho_{\omega}^{2}}\right),n=0,1,2...,\;
l=0,\pm1,\pm2,...,
\end{equation}
$$ Y_{l}(\theta)= \frac{1}{\sqrt{2\pi}}e^{il\theta},\;\;\;
R_{k_{z}}(z)=\frac{1}{\sqrt{V}}e^{ik_{z}z}.$$ The radial functions
$\Psi_{nl}(\rho)$ comprise the generalized Laguerre polynomials
$L_{n}^{|l|}$. The value $\rho_{\omega}=\sqrt{\hbar/m_{e}\omega}$
serves as a characteristic scale in our problem. The function
$R_{k_{z}}(z)$ corresponds to a plane wave with $k_{z}$ being the
longitudinal wave number and normalization is based on the
property $\delta(0)/V\rightarrow 1$ as $V\rightarrow\infty$. The
functions (2.6) form a complete orthonormal set. The energy
corresponding to the state $\chi_{nlk_{z}}$ is given by
\begin{equation}
E^{(0)}_{{nlk_{z}}}=\hbar\omega(2n+|l|+1)+\frac{\hbar^{2}k_{z}^{2}}{2
m_{e}}.
\end{equation}
 It is important that the whole
energy may be separated into two conserving constituents: the
lateral and longitudinal ones, and the lateral spectrum is
equidistant. This property holds true as long as Hamiltonian
depends on $z$ only by the way of derivatives
$\partial^{n}/\partial z^{n}$. The longitudinal component of the
wave vector, $k_{z}$, remains constant during the motion and is
convenient to be considered as a parameter. At a given $k_{z}$ the
eigenfunctions (2.6) form a complete set in the corresponding
subspace and may be chosen as a basis for seeking the approximate
solutions of the eigenvalue problem
\begin{equation}
\hat{H}\Phi=E\Phi.
\end{equation}
The Hamiltonian (2.2) has a $2\times2$-matrix structure, so its
basis eigenfunctions must be of a two-component vector form. Such
two-component functions form the fundamental system of the matrix
Hamiltonian
\begin{equation}
\left(
\begin{array}{cc} \hat{H}_{0} & 0 \\ 0 & \hat{H}_{0}
\end{array}\right)
\end{equation}
and read:
\begin{equation}
\chi_{\uparrow,\;nlk_{z}}= \left(
\begin{array}{cc}
\chi_{nlk_{z}}  \\ 0
\end{array}\right),\;\;\;
\\ \chi_{\downarrow,\;nlk_{z}}= \left(
\begin{array}{cc}
0  \\ \chi_{nlk_{z}}
\end{array}\right).
\end{equation}

The diagonal $\langle
\chi_{(\uparrow,\downarrow),\;nlk_{z}}|V_{SO}|
\chi_{(\uparrow,\downarrow),\;mkk_{z}}\rangle$ and
off-dia\-go\-nal $\langle
\chi_{(\uparrow,\downarrow),\;nlk_{z}}|V_{SO}|
\chi_{(\downarrow,\uparrow),\;mkk_{z}}\rangle$ matrix elements as
well as matrix elements which appear in Section III may be easily
calculated. The convolutions of two-component vectors with
$2\times2$ matrices were performed according to the usual matrix
multiplication rules. The integrals needed to be taken in order to
calculate the matrix elements may be written as a product of three
integrals over $\rho,\;\theta$ and $z$.
%\widetext%
%\begin{equation}
%\int \rho d\rho \; d\theta \; dz
%f(\rho,\theta,z)=\int_{0}^{\infty} \rho d\rho f_{\rho}(\rho)
%\int_{0}^{2\pi} d\theta f_{\theta}(\theta) \int_{-\infty}^{\infty}
%dz f_{z}(z).
%\end{equation}
The integrals over $\rho$ were taken according to the formula
\cite{Prudnikov}: \\%
\vspace{\baselineskip}
\begin{widetext}
\begin{displaymath}
\int_{0}^{\infty}x^{\alpha}\rho_{\omega}^{2}
\chi_{mk}(x)\chi_{nl}(x)dx=
\frac{\Gamma(1+\alpha+(l+k)/2)}{\sqrt{m!\,n!\,(m+|k|)!\,(n+|l|)!}}\frac{\Gamma(1+n+l
)} {\Gamma(1+l)}\frac{\Gamma((k-l)/2-
\alpha+m)}{\Gamma(2+(l-k)/2+\alpha)}\times
\end{displaymath}
\begin{equation}
{}_{3}F_{2}[-n,2+\alpha+(l+k)/2,2+\alpha+(k-l)/2;1+k,2+\alpha-m+(k-l)/2;1].
\end{equation}
\end{widetext}
In concordance with the normalization rules for $Y_{l}(\theta)$
\cite{Landau} the integrals over $\theta$ and $z$ are simple
Fourier transformations. For instance, what follows is a typical
integral one meets in the calculation of the matrix elements from
Section III:
\begin{equation}
\int_{0}^{2\pi}
\cos^{2}(\frac{N\theta}{2})e^{i(k-l)\theta}d\theta=\frac{\pi}{2}(2\delta_{k,l}
+\delta_{k-l,N}+\delta_{k-l,-N}).
\end{equation}

In numerical calculations throughout this paper we used 30 up- and
\begin{figure}[h!]
\centering %
\scalebox{1}[1]{\includegraphics[289,98][517,478]{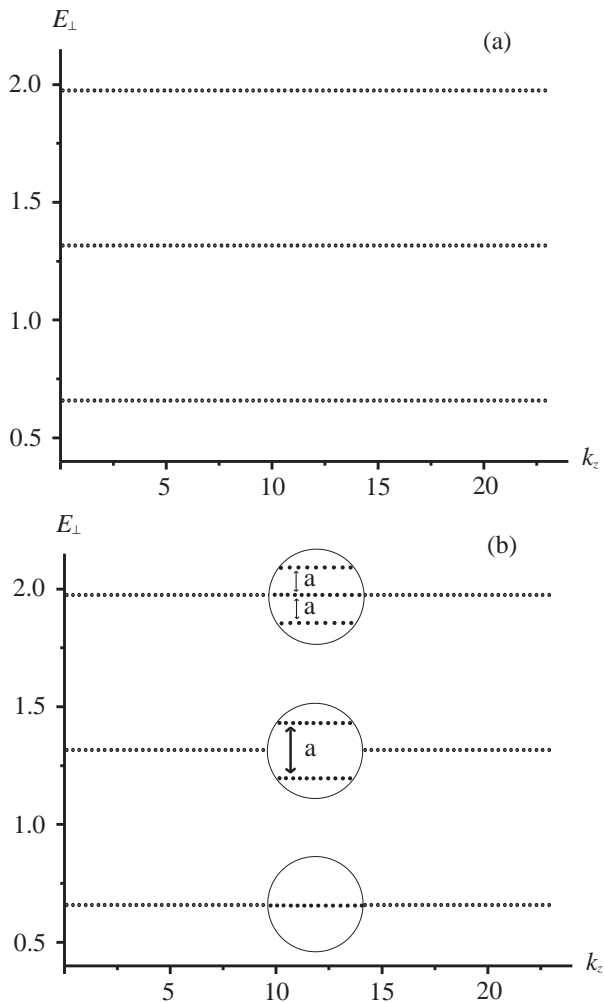}}%
%\scalebox{1}[1]{\includegraphics[190,147][385,543]{files/fig1.eps}}%
\caption{\label{f_1} (a) The spectrum of $\hat{H}_{0}$; (b) the
spectrum of $\hat{H}$, $\mu = 1$. Energy $E$ is measured in eV,
$k_{z}$ is measured in units of $10^{6} cm^{-1}$, $a$ = 8.47$\cdot
10^{-6}$~eV.}
\end{figure}
30 down-basis functions. This allowed us to obtain 60 energy
levels with a difference from the exact values only in the 9th
significant digit.
%It should be noted that this numerical scheme is much
%simpler than the one exploited in the work \cite{Moroz}.
The value of $\omega$ was chosen to be
$10^{15}\;\mathrm{sec^{-1}}$ and $m_{e}^{*}$ was put equal to
$0.05m_{e}$.

The Fig.~\ref{f_1} shows several first lateral energy levels of
the Hamiltonians $\hat{H}_{0}$ and $\hat{H}$. The lateral energies
pertaining to $\hat{H}_{0}$ are degenerate with the multiplicity
$2(2n+|l|+1)$. Thus, every $q$-th eigenvalue (counting from the
bottom of the quantum well) of $\hat{H}_{0}$ is $2q$ times
degenerate and splits into q branches in transition to $\hat{H}$.
Therefore, $\hat{V}_{SO}$ leaves 2-fold spin degeneracy of every
energy level.
%%%%%%%%%%%%%%%%%%%%%%%%%%%%%%%%%%%%%%%%%%%%%%%%%%%%%%%%%%%%%%%%%%%%%%%%%%
\section{Perturbed axially symmetric confining potential}
\subsection{Non-electromagnetic perturbation}
The perturbation we will deal with here and in the subsequent part
of this section is
\begin{equation}
\hat{V}_{pert}=A\frac{m_{e}\omega^{2}}{2}\rho^{2}\cos^{2}(\frac{N\theta}{2}),
\end{equation}
\begin{equation}
A\geq 0,\;\; N\in 0,1,2,...
\end{equation}

\begin{figure}[t!]
\centering %
\scalebox{1}[1]{\includegraphics[333,202][555,390]{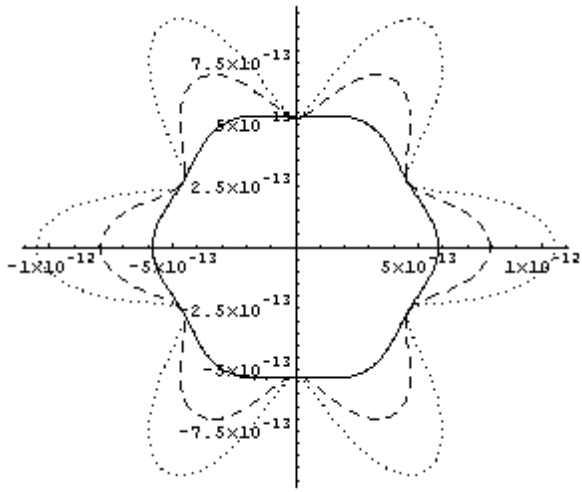}}%
%\scalebox{1}[1]{\includegraphics[190,147][385,543]{files/fig2.eps}}%
\caption{\label{f_2} The polar plot of the potential
$\hat{V}_{conf} + \hat{V}_{pert}$, $N=6$, $\rho=\rho_{\omega}$:
dotted line -- $A = 1.0$, dashed line -- $A = 0.5$, solid line --
$A = 0.1$.}
\end{figure}

The dimensionless constant parameters $A$ and $N$ define the
amplitude and the number of `lobes' of the perturbation,
respectively (see Fig.~\ref{f_2}). At $N=2$ our problem resembles
the 2D type of the problem solved by Migdal \cite{Migdal}. In this
part we investigate the effect which $\hat{V}_{pert}$ has on the
spectrum considering it to be of non-electromagnetic nature
($\hat{V}_{pert}^{n-el}$). Under the non-electromagnetic nature of
the perturbation we imply that it does not affect $\varphi$ and
$\vec{E}$ in the Hamiltonian (2.4), but is simply added to the
right-hand side of (2.4). On the other hand, the electromagnetic
perturbation $\hat{V}_{pert}^{el}$, which is regarded in the next
subsection, does influence $\varphi$ and $\vec{E}$.

\begin{figure}[t!]
\centering%
\scalebox{1}[1]{\includegraphics[228,186][456,712]{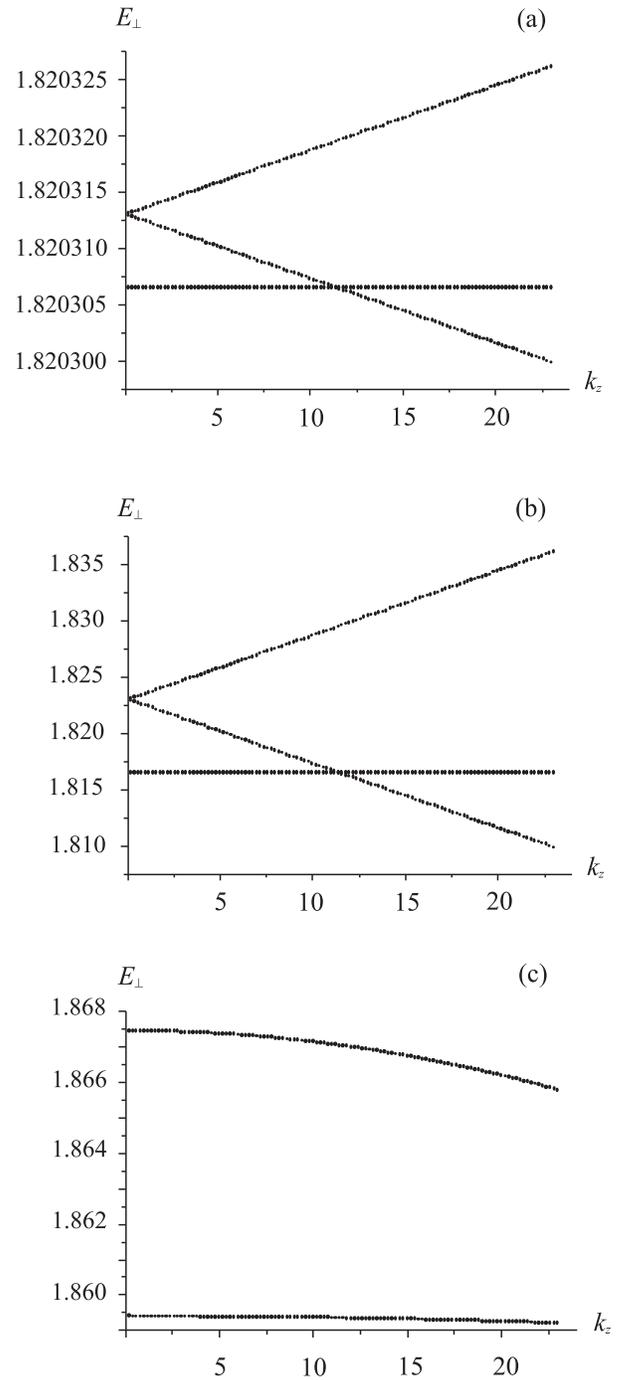}}%
\caption{\label{f_3} The spin splitting originated by
non-electromagnetic perturbation $\hat{V}^{n-el}_{pert}$. Energy
$E$ is measured in eV, $k_{z}$ is measured in units of $10^{6}
cm^{-1}$. (a) $A = 1$, $\mu=1$, $N=3$, $d = 1, 2$; (b) $A = 1$,
$\mu=10^{3}$, $N=3$, $d = 1, 2$; (c) $A = 1$, $\mu=10^{3}$, $N=4$,
$d = 1, 2$.}
\end{figure}

The case of the non-electromagnetic perturbation is especially
interesting because at even $N$ $\hat{V}_{pert}^{n-el}$ leaves the
inversion symmetry ($\theta\rightarrow\theta+\pi,z\rightarrow-z$)
of the Hamiltonian intact, although breaks its axial symmetry.
This should have a non-trivial effect on the spin splitting in the
energy spectra.
%
%This left symmetry may result in several or even all energy levels
%being still twice-degenerate.
%%%%%%%%%%%%%%%%%%%%%%%%%%%
%%%%%%%%%%%%%%%%%%%%%%%%%%%
The increased (in comparison with the free space value $\mu=1$)
value of $\mu$, equal to $10^{3}$, approximately corresponds to
that of the constant $\beta$ from the work \cite{Moroz} and may
not only allow to draw correlations with the results of the
latter, but can also help to elucidate the role of SO interaction
in the formation of energy band structure. The Fig.~\ref{f_3}
demonstrates spin splitting for various values of the parameters
$N$, $A$, and $\mu$. At even values of $N$ no spin splitting is
predicted for any energy level. This may be explained by higher
symmetry of the potential in this case. At odd $N$ the spin
splitting follows a rather peculiar pattern: with the growth of N
the number of non-perturbed energy level (if the ground state has
number zero) at which spin splitting occurs for the first time
enhances. The Table 1 illustrates this behavior.

\bigskip
Table 1. Energy levels of the Hamiltonian H which undergo
splitting in the case of non-electromagnetic perturbation
$\hat{V}_{pert}^{n-el}$. The ground state has number zero.
\begin{center}
\scalebox{0.98}[1]{\begin{tabular}{|c|c|c|c|c|c|c|c|} \hline $N$ &
1 & 2 & 3 & 4 & 5 & 6 & 7 \\ \hline Level number & all & none &
2,3,9,... & none & 4,6,... & none & 7,10,... \\ \hline
\end{tabular}}
\end{center}
\bigskip

Very important for practical reasons is the question of how spin
splitting depends on the amplitude of perturbation $A$. The Table
2 shows that the magnitude of spin splitting decreases very
rapidly as $A$ gets smaller, so that minute perturbations give
almost indiscernible splitting. The rise of the contribution of SO
interaction (the increase of $\mu$) does not lead to any
fundamental changes in the energy spectra, but increases almost
proportionally to $\mu$ the values of spin splitting. For example,
at $A=1$, $N=3$, $\mu=10^{3}$ and
$k_{z}=2\cdot10^{7}\mathrm{cm}^{-1}$ spin splittings
$|\triangle_{d}|$, where $d$ is the number of the unperturbed
energy level, for $d=2$ and $d=3$ are $2.32\cdot10^{-2}$ eV and
$3.07\cdot10^{-2}$ eV, respectively.
%%%%%%%%%%%%%%%%%%%%%%%%%%%%%%%%%%%%%%%%%%%%%%%%%%%%%%%%%%%%%%%%%%%%%%%%%%
\subsection{Electromagnetic perturbation}
\begin{figure}[b!]
\centering%
\scalebox{1}[1]{\includegraphics[239,253][453,576]{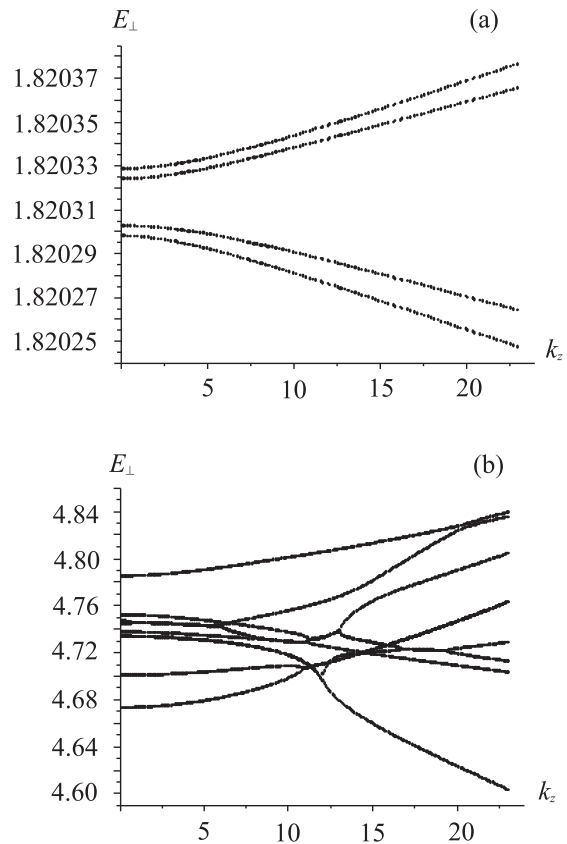}}%
\caption{\label{f_4} The spin splitting originated by
electromagnetic perturbation $\hat{V}^{el}_{pert}$. Energy $E$ is
measured in eV, $k_{z}$ is measured in units of $10^{6} cm^{-1}$.
(a) $A = 1$, $\mu = 1$, $N = 3$, $d = 1, 2$; (b) more complex
behavior of high-lying levels, $A = 1$, $\mu = 10^{3}$, $N = 4$,
$d = 21,\ldots, 25$.}
\end{figure}

Treated as being of electromagnetic nature, the perturbation
$\hat{V}_{pert}^{el}$, which formally has the form (3.1), results
in the following Hamiltonian:
\begin{equation}
\hat{H}'=\hat{H}+\triangle \hat{H}.
\end{equation}
Here \widetext
$$\triangle \hat{H}=-\frac{i\hbar^{2}\omega^{2}}{4m_{e}c^{2}}
\left\{A\rho\cos^{2}(\frac{N\theta}{2})\left(
\begin{array}{cc}
0 & -i e^{-i\theta} \\ i e^{i\theta} & 0
\end{array}\right)
\frac{\partial}{\partial
z}+A\frac{\hbar^{2}\omega^{2}}{8m_{e}c^{2}}\left[1
-(\frac{N^{2}}{4}-1)\cos(N\theta)\right]+\right.$$
\begin{equation}
\;\;\;\;\;\;\;\;\;\;\;\;\;\; \left.
\frac{AN}{2}\rho\cos(\frac{N\theta}{2})\sin(\frac{N\theta}{2})
\left(
\begin{array}{cc}
0 & e^{-i\theta} \\ e^{i\theta} & 0
\end{array}\right)\frac{\partial}{\partial
z} -\frac{AN}{2}\rho\cos(\frac{N\theta}{2})\sin(\frac{N\theta}{2})
\sigma_{z}\frac{\partial}{\partial \rho} -
A\cos^{2}(\frac{N\theta}{2})\sigma_{z}
\frac{\partial}{\partial\theta}\right\}.
\end{equation}
\endwidetext
The matrix elements of $\triangle \hat{H}$ in space of the
functions (2.13) can also be explicitly calculated (see Eqns.
(2.13)-(2.15)). The Fig.~\ref{f_4} demonstrates the spin splitting
originated by $\hat{V}_{pert}^{el}$. Now all the energy levels are
non-degenerate for both even and odd values of $N$. The magnitude
of spin splitting goes down with the decrease of $A$ approximately
with the same rate as it does in the case of non-electromagnetic
perturbation (see Table 2). It is clear from the Table 2 that as
$A$ decreases, the difference between non-electromagnetic and
electromagnetic cases gradually vanishes.%
\vspace{\baselineskip}

\bigskip
Table 2. The value of spin splitting $\triangle_{d}$ (eV) at
various amplitudes $A$ of the perturbation $\hat{V}_{pert}$.
$N=3$, $\mu=1$, $k_{z}=2\cdot10^{7} \; \mathrm{cm}^{-1}$.%
\begin{center}
\scalebox{0.97}[1]{\begin{tabular}{|c|c|c|c|c|c|} \hline
\multicolumn{2}{|c|}{$A$} & $1$ & $10^{-1}$ & $10^{-2}$ & $10^{-3}$ \\
\hline
 $\hat{V}_{pert}^{n-el}$ & $|\triangle_{2}|$ & $2.29\cdot10^{-5}$
 & $4.16\cdot10^{-6}$ & $4.06\cdot10^{-7}$ & $4.05\cdot10^{-8}$ \\
\hline
 $\hat{V}_{pert}^{n-el}$ & $|\triangle_{3}|$ & $3.04\cdot10^{-5}$
 & $6.16\cdot10^{-6}$ & $6.09\cdot10^{-7}$ & $6.07\cdot10^{-8}$ \\
\hline
 $\hat{V}_{pert}^{el}$ & $|\triangle_{2}|$ & $9.63\cdot10^{-6}$
 & $2.87\cdot10^{-6}$ & $4.04\cdot10^{-7}$ & $4.05\cdot10^{-8}$ \\
\hline
 $\hat{V}_{pert}^{el}$ & $|\triangle_{3}|$ & $1.75\cdot10^{-5}$
 & $4.65\cdot10^{-6}$ & $6.08\cdot10^{-7}$ & $6.07\cdot10^{-8}$ \\
\hline
\end{tabular}}
\end{center}
\bigskip
%%%%%%%%%%%%%%%%%%%%%%%%%%%%%%%%%%%%%%%%%%%%%%%%%%%%%%%%%%%%%%%%%%%%%%%%%%
\section{Conclusion}
In this work we studied theoretically a 3D gas of non-interacting
electrons which was transformed into a Q1DEG by means of 2D
confining potential. We have solved the Pauli equation numerically
for this system with perturbed axially-symmetric, parabolic
confining potential. The non-perturbed axially symmetric potential
$\hat{V}_{conf}$ was assumed to be created by electromagnetic
field with the scalar potential
$\varphi=m_{e}\omega^{2}\rho^{2}/2e$. Two different types of
perturbations capable of lifting the spin degeneracy are
considered: of electromagnetic and non-electromagnetic nature. We
have found that these two types of perturbation have different
effects on the energy spectra. While the electromagnetic
perturbation splits all the levels, the non-electromagnetic one in
most cases (besides the case with $N=1$) leaves the spin
degeneracy of some levels. We have found that for large
perturbations ($A=1$) the value of spin splitting is large enough
to be detected in the experiment ($10^{-5}$ eV and higher). As $A$
decreases, spin splitting also goes down linearly or even faster.
Therefore, if practically created axially-symmetric potentials
only slightly deviate from the axial symmetry, spin splitting
originated by discussed in this work mechanism is unlikely to be
observed. However, if a confining potential possesses only rough
axial symmetry, the additional splitting of the energy levels may
appear.

%%%%%%%%%%%%%%%%%%%%%%%%%%%%%%%%%%%%%%%%%%%%%%%%%%%%%%%%%%%%%%%%%%%%%%%%%%
\bigskip
\begin{acknowledgments}
We are very grateful to S.A.~Derevyanko, Yu.P.~Ste\-pa\-nov\-sky,
and V.A.~Yam\-pol'\-skii for many fruitful discussions.
\end{acknowledgments}

\end{document}